\documentclass[prd, preprint, 11pt]{revtex4-1}

\usepackage{amsmath}
\usepackage{amssymb}
\usepackage{setspace}
\usepackage{graphicx}
\usepackage{natbib}
\usepackage{float}

\begin{document}

\begin{center}
 { \large {\bf The problem of time and the problem of quantum measurement}

\vskip 0.2 in


{\large{\bf Tejinder  P. Singh}} 

{\it Tata Institute of Fundamental Research,}
{\it Homi Bhabha Road, Mumbai 400 005, India}
\tt email: tpsingh@tifr.res.in}\\
\medskip

\end{center}

\centerline{\bf ABSTRACT}
\smallskip

\noindent Quantum theory depends on an external classical time, and there ought to exist an equivalent reformulation of the theory which does not depend on such a time. The demand for the existence of such a reformulation suggests that quantum theory is an approximation to a stochastic non-linear theory. The stochastic 
non-linearity provides a dynamical explanation for the collapse of the wave-function during a quantum measurement. Hence the problem of time and the measurement problem are related to each other: the search for a solution for the former problem naturally implies a solution for the latter problem.

\vskip 2 in

\centerline{\it Based on a talk given at the conference Quantum Malta 2012: Fundamental Problems in Quantum Physics}
\centerline{\it  University of Malta, Malta, April 24-27, 2012}
\vskip 0.2 in
\centerline{\it Submitted to the volume {\bf The Forgotten Present} (running title), Thomas Filk and Albrecht von M\"{u}ller (Editors)}

\vskip 1.2 in

\centerline{\it Dedicated to Malala Yousafzai, for her extraordinary courage and support for the cause of education and knowledge}
\newpage
\setstretch{1.1}
\noindent

\section{Why remove classical time from quantum theory?}
\noindent Dynamical evolution in quantum theory is described by the Schr\"{o}dinger equation. The time parameter which is used for describing this evolution is part of a classical spacetime. By classical spacetime we mean both the underlying spacetime manifold, as well as the gravitational field [equivalently the metric] which resides on it. As we know, the gravitational field is determined by the distribution of classical matter according to the laws of the general theory of relativity. What is perhaps not so well appreciated is that, in accordance with the Einstein hole argument, a physical meaning cannot be attached to the points of the underlying manifold unless a dynamically determined metric tensor field resides on it ~\cite{Christian:98, Singh:2009}. Thus one can reasonably assert that classical spacetime, and hence also the time parameter used to describe evolution in quantum theory, is determined by classical bodies and fields. Now, the dynamics of classical objects is itself a limiting case of quantum dynamics. We see here the circularity of time in quantum theory. Quantum theory depends on classical time. But classical time is well-defined only after one considers the classical limit of quantum theory (Fig. 1).

\begin{figure} [ht]
  \centerline{\includegraphics{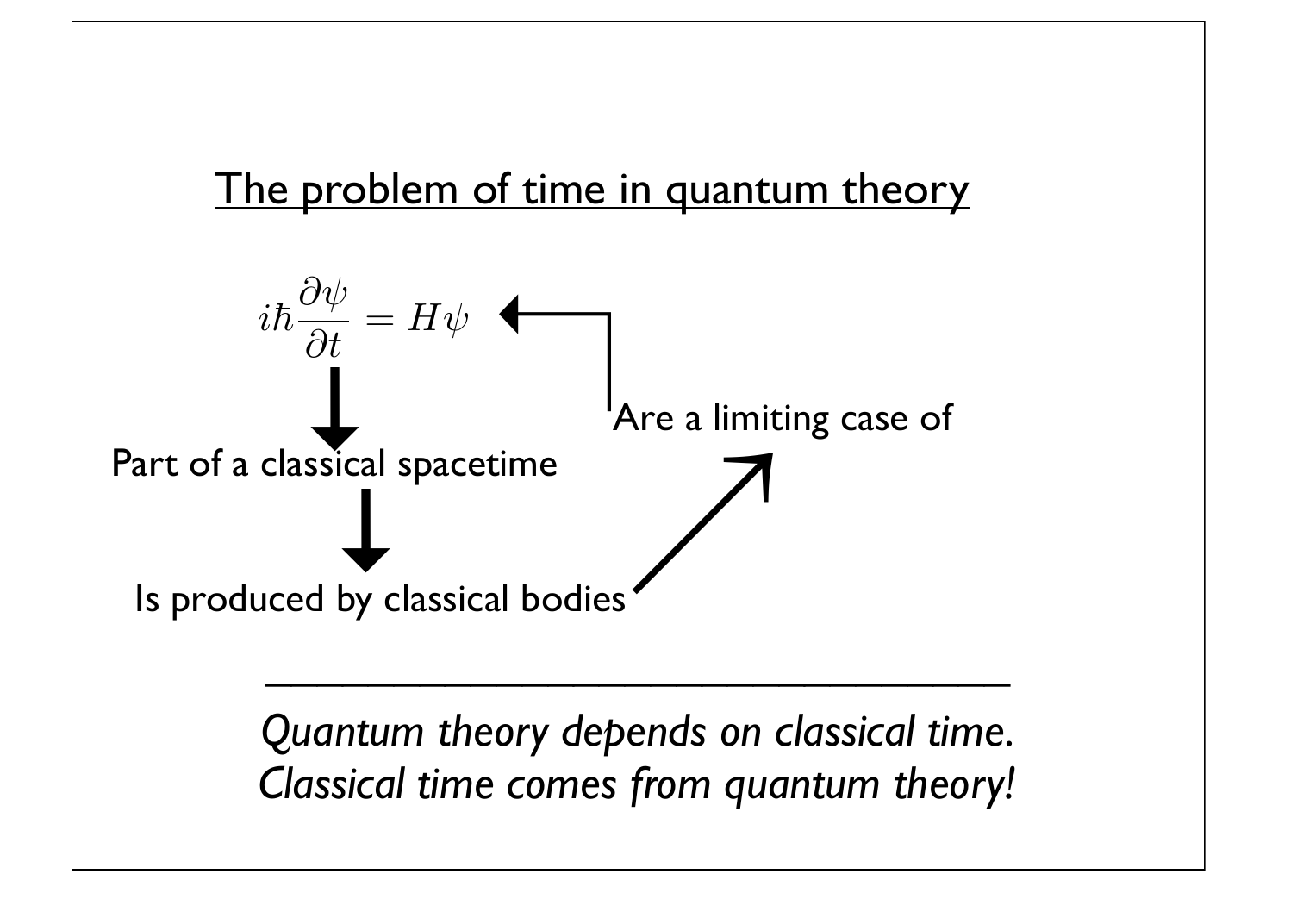}}
\caption{The circularity of time in quantum theory}  
\end{figure}%

We hence conclude that there should exist an equivalent new formulation of quantum theory which does not depend on classical time. We have argued elsewhere that such a new formulation is a limiting case of a stochastic  non-linear theory. The non-linearity, which has to do with gravity,  becomes significant in the approach to the Planck mass/energy scale and possibly plays a role in explaining the collapse of the wave-function during a quantum measurement ~\cite{Singh:2006, Singh:2009}.

How should one go about constructing such a reformulation, which we will call Generalized Quantum Dynamics [GQD]? One is foregoing classical time, and along with it, the point structure of a spacetime manifold. A natural possibility is to replace the original spacetime by a non-commutative spacetime. Such a spacetime, and its associated dynamics, called Non-commutative Special Relativity [NSR], was proposed by us in a recent work ~\cite{Lochan-Singh:2011}. In NSR, evolution is described via a `proper time' constructed from taking the Trace over the non-commutative spacetime metric.

As will be described in the next section, a GQD is arrived at by constructing the equilibrium statistical thermodynamics of the underlying non-commutative special relativity ~\cite{Lochan:2012}. Section IV then sketches ongoing work on how one possibly recovers classical spacetime and classical matter fields, from considerations of statistical fluctuations around a GQD. This work, when complete, would be central to achieving a fundamental understanding of why superpositions of position states are absent in the macroscopic, classical world (Fig. 2).

\begin{figure} [ht]
  \centerline{\includegraphics{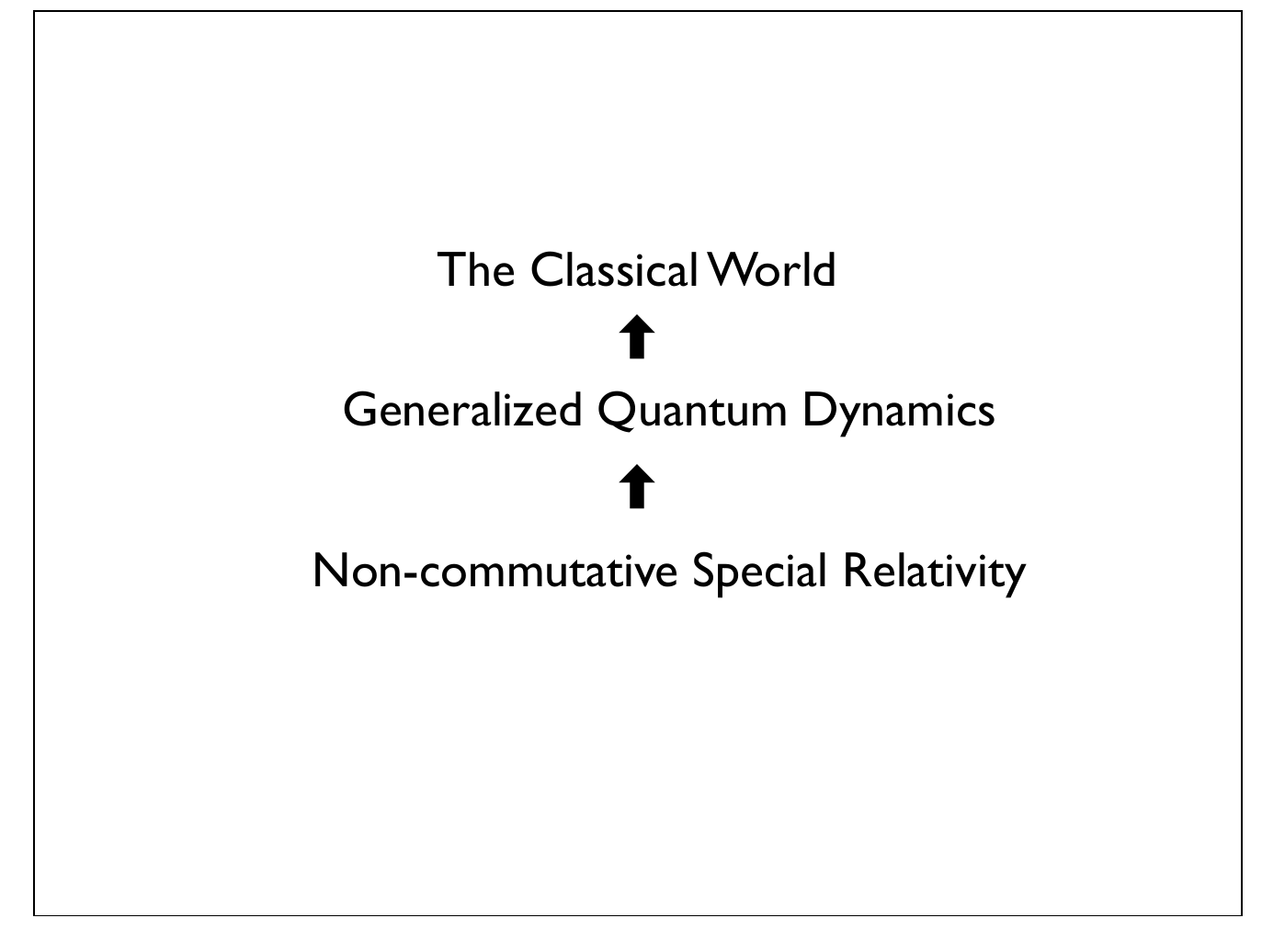}}
\caption{From a non-commutative spacetime to the classical world, via GQD}  
\end{figure}%

One notices that in the transition from a GQD to the classical world, there is no sign of ordinary quantum theory [which depends on classical time]! That recovery must take place separately, and that is where the connection of the time problem with the measurement problem emerges. In Fig. 2, by classical world is meant a universe which is {\it dominated} by classical matter fields. Only when such a dominance is given, can one talk of the existence of a classical spacetime; otherwise the Einstein hole argument will again come into play and forbid the occurrence of the ordinary spacetime manifold. However, not all matter is classical; there is a sprinkling of `quantum' fields, whose dynamics must be derived from first principles, given a classical time.

This is what is achieved by the theory of Trace Dynamics ~\cite{Adler:04, Adler:94, Adler-Millard:1996, RMP:2012} which is the classical dynamics of non-commuting matrices on a background classical spacetime. The equilibrium statistical thermodynamics of this matrix dynamics is shown to be the ordinary quantum theory.  Statistical fluctuations around equilibrium are shown to lead to non-linear modifications of the quantum theory, and this non-linearity is responsible for collapse of the wave-function during a quantum measurement (Fig. 3). In the limit when the non-linearity becomes strongly dominant, the non-linear theory reduces to classical mechanics.

\begin{figure} [ht]
  \centerline{\includegraphics{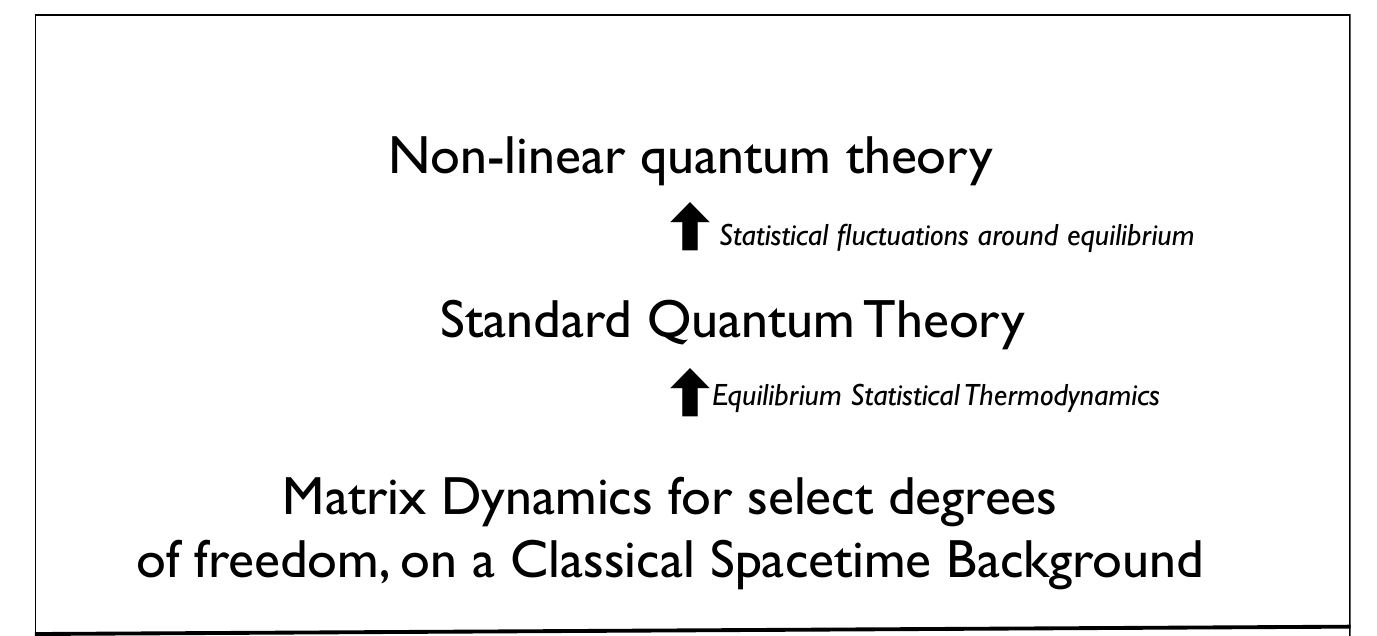}}
\caption{From Trace Dynamics to a nonlinear theory, via standard quantum theory}  
\end{figure}%

The connection between the problem of time and the problem of measurement is the following. In our opinion, Trace Dynamics should perhaps not be treated as a stand-alone theory. Because it gives a matrix (equivalently operator) status to matter degrees of freedom, while retaining a point-like structure for spacetime. This will again run into the kind of difficulties implied by the Einstein hole argument: a non-commutative nature for matter degrees is not consistent with a commutative nature for spacetime degrees, unless a dominant classical matter background is available. Thus, a logical starting point for Fig. 3 is to place it at the top of Fig. 2. First one starts from a non-commutative special relativity and derives a GQD, and from there the classical world with a classical time.  On this classical world one considers the matrix dynamics for select degrees of freedom (which are sub-dominant and not classical), and this eventually leads to a non-linear quantum theory. The physics which solves the problem of time in quantum theory is strongly correlated with the physics that solves the measurement problem in quantum theory (Fig. 4). 

Fig. 4 captures the philosophy of our approach, and the essence of this article. One starts from an NSR and arrives at a GQD. This is described in Section II. The transition from a GQD to the classical world is discussed in Section IV (the logical place would be Section III, but this work is as yet incomplete, and hence its discussion is left till the end). The derivation of ordinary quantum theory and the solution of the quantum measurement problem is discussed in Section III.

\begin{figure} [ht]
  \centerline{\includegraphics{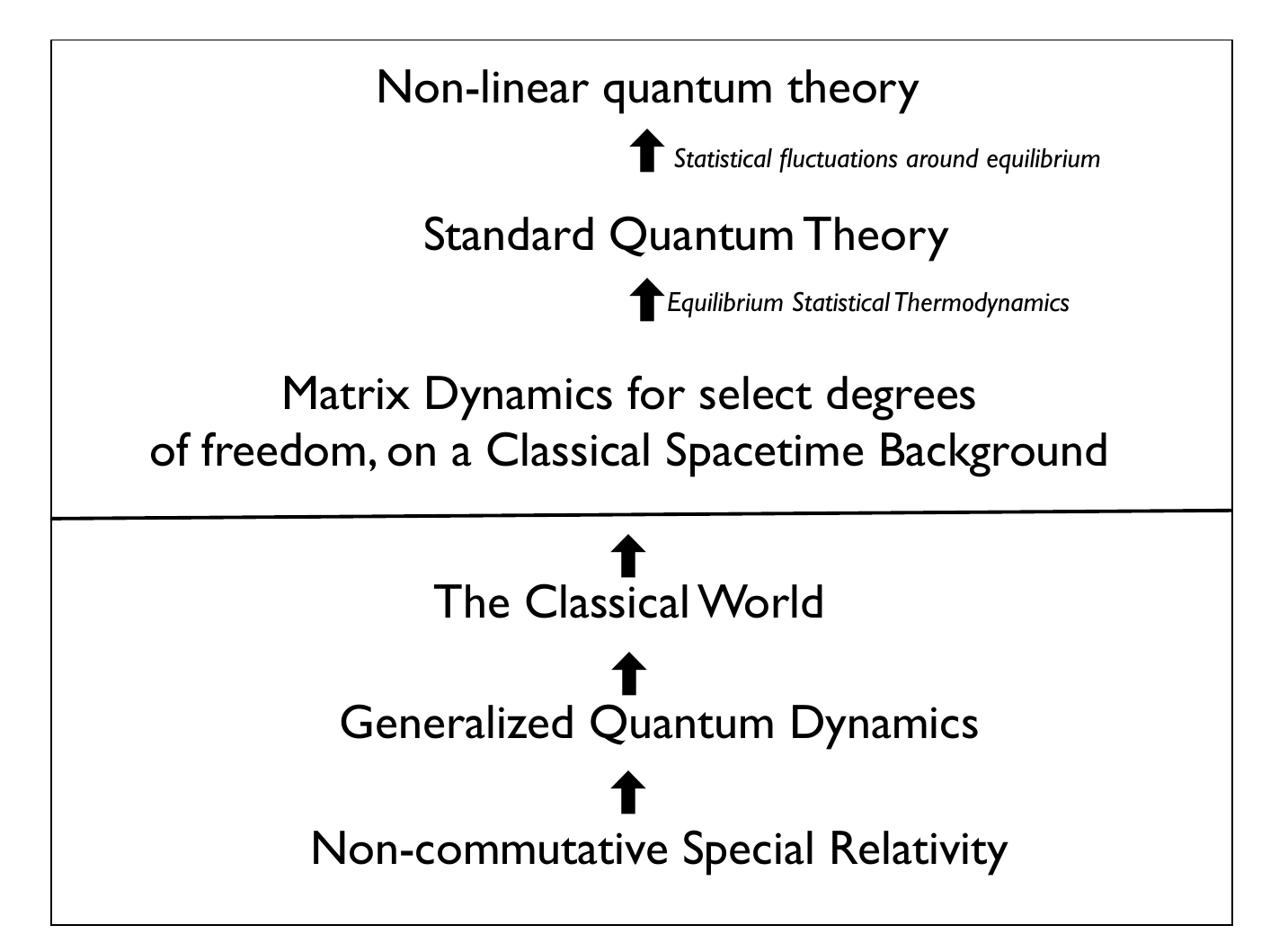}}
\caption{Solving the problem of time, and the problem of quantum measurement}  
\end{figure}%

\section{A generalized quantum dynamics}
The mathematical formulation leading up to a GQD ~\cite{Lochan:2012} is strongly motivated by and based on the theory of Trace Dynamics developed by Stephen Adler and collaborators ~\cite{Adler:04}. The new added element is the assumption of a non-commutative spacetime with operator (equivalently matrix) coordinates $(\hat{t}, \hat{x}, \hat{y}, \hat{z})$, for which a proper time is defined by taking a trace over a line-element:
\begin{equation}
ds^{2} = Trd\hat{s}^2\equiv Tr[d\hat{t}^2 - d\hat{x}^2 - d\hat{y}^2 - d\hat{z}^2].
\end{equation}
This line element is invariant under coordinate transformations of the non-commuting coordinates, with their commutation relations being completely arbitrary. Fermionic / Bosonic matter degrees of freedom, described by non-commuting matrices, live on this spacetime, and are respectively characterized by whether they belong to odd / even  sector of the graded Grassmann algebra. A classical dynamics of these non-commuting matrix degrees of freedom $\hat{q}^i$ can be constructed to describe evolution with respect to the proper time $s$: we call this a non-commutative special relativity [NSR]. Thus as in special relativity, a `particle' is assigned a set of four coordinates $(\hat{t}, \hat{x}, \hat{y}, \hat{z})$, a four velocity is defined by taking their derivative with respect to the proper time, and a canonically conjugate four momentum $\hat{p}^{i}$ is defined by taking the `trace derivative' of the Trace Lagrangian (trace of a polynomial function of coordinates and velocities) with respect to the four velocity. From the Trace Lagrangian, one derives Lagrange equations of motion, a Trace Hamiltonian, and Hamilton's equations, as in ordinary mechanics ~\cite{Lochan-Singh:2011}.

The central feature of this matrix classical dynamics, which makes it different from point particle classical dynamics, is that it possesses a novel conserved charge:
\begin{equation}
 \hat{Q} = \sum_{r\in B}[\hat{q}_r,\hat{p}_r] -\sum_{r\in F}\{\hat{q}_r,\hat{p}_r\},
\end{equation}
where the commutators are for bosonic degrees of freedom, and anticommutators are for fermionic degrees. 
We note that the commutators/anti-commutators also include pairs such as $[\hat{E}^i,\hat{t}^i]$ and $\{\hat{E}^i,\hat{t}^i\}$, where $\hat{E}^i$ is the energy variable canonically conjugate to $\hat{t}^{i}$. 
This conserved charge $\hat{Q}$, which has the dimensions of action, is a consequence of the global unitary invariance of the Lagrangian and the Hamiltonian. It would be trivially zero in the case of point-particle mechanics, but that is not the case here, and its existence is all the more remarkable, because the individual $q-q$, $q-p$, and $p-p$ commutators / anti-commutators are non-zero and completely arbitrary. The existence of this charge plays a central role in the emergence of quantum theory from this underlying level, as we will see shortly.

This matrix dynamics on a non-commutative `flat' space-time is according to us the fundamental dynamics, its symmetries being invariance of the operator spacetime metric under Lorentz transformations, and the global unitary invariance of the Lagrangian. 

However this is not the dynamics we observe in our laboratory experiments. Hence one proposes that this dynamics must be coarse-grained over, much the same way that coarse graining over the microscopic degrees of freedom reproduces the statistical thermodynamics of macroscopic systems. Thus we shall develop the statistical thermodynamics of the above classical matrix dynamics, employing entirely conventional methods and techniques of equibrium statistical mechanics. The classical matrices are analogous to the atoms of a gas, and the coarse-graining is anaologous to constructing the thermodynamics of the gas, leading to its approximate macroscopic thermodynamic description. It is remarkable that the thermodynamics of this matrix dynamics will be the sought for GQD, which is a precursor to quantum theory, and in that sense quantum theory is an emergent phenomenon. 

One starts by showing that a measure $d\mu$ can be defined in the phase space of the matrix degrees of freedom, and Liouville's theorem holds,  demonstrating the conservation of phase space volume. A probability density distribution $\rho(H,T;\hat{Q},\lambda)$ is defined in the phase space, where the `temperature' $T$ and the matrix $\lambda$ are respectively the Lagrange multipliers introduced to respect the conservation of the Hamiltonian and the charge $\hat{Q}$. A canonical ensemble is constructed and an equilibrium distribution is arrived at by maximizing the entropy 
\begin{equation}
 S = - \int d \mu \rho \log \rho
\end{equation}
subject to the conservation constraints. As anticipated, the equilibrium distribution is given by
\begin{equation}
\rho = Z^{-1} \exp(Tr\lambda \hat{Q} - HT) 
\end{equation}
with $Z$ being the partition function. An important result which can be proved is that the canonical ensemble average of $\hat{Q}$ is of the form
\begin{equation}
\langle\hat{Q}\rangle_{AV} = i_{eff}\hbar
\end{equation}
where $\hbar$ is a real positive constant of dimensions of action, and $i_{eff} = diag(i,-i,i,-i,...,i,-i)$ such that $Tri_{eff}=0$.

Now, the phase space measure, as well as the canonical average of an observable ${\cal O}$ given by
\begin{equation}
\langle {\cal O} \rangle_{AV} = \int d\mu \rho {\cal O}
\end{equation}
are invariant under constant shifts of dynamial variables in phase space. This leads to an important Ward identity for a polynomial function $W(z)$ of the dynamical variables $z$ in phase space. 

Under the assumptions that $T$ is identified with the Planck scale, and we work much below that scale, and secondly that in the Ward identity the conserved charge $\hat{Q}$ can be replaced by its canonical average
$i_{eff}\hbar$ the Ward identity simplifies greatly, to the following 
\begin{equation}
\langle {\cal D} z_{eff}\rangle_{AV}=0; \quad  {\cal D}z_{r eff} =i_{eff}[W_{eff}, z_{r eff}]-\hslash\sum_s\omega_{rs}\left(\frac{\delta {\bf W}}{\delta z_s}\right)_{eff}. \label{ModWard}
\end{equation}

This equation contains the essence of the sought for GQD! Here, $z_{eff}$ is that matrix component of the matrix dynamical variable $z$ which commutes with $i_{eff}$. Different choices of the polynomial $W$ lead
to different important results which contain the mathematical essence of GQD.

If $W$ is chosen to be the operator Hamiltonian $H$, this Ward identity becomes the Heisenberg equations of motion
\begin{equation}
\langle {\cal D} z_{eff}\rangle_{AV}=0; \quad  {\cal D}z_{r eff} =i_{eff}[H_{eff}, z_{r eff}]-\hslash\dot{z}_{r eff}. \label{ModWard1}
\end{equation}
A dot denotes derivative with respect to the proper time $s$. We recall that the operator time $\hat{t}$ is one of the dynamical variables $z$. 

Next, if we choose $W=\sigma_vz_v$ we get
\begin{equation}
 i_{eff}{\cal D}z_{r eff} =[z_{r eff}, \sigma_v z_{v eff}]-i_{eff}\hslash\omega_{rv}\sigma_v \label{ModWard2}
\end{equation}
which gives the emergent canonical commutation rules for the bosonic and fermionic degrees of freedom. Thus we obtain, what we call effective canonical commutators of the canonically averaged matter degrees of freedom.
For a bosonic pair
\begin{equation}
 [q^{\mu},q'^{\nu}]=0; \quad  [q^{\mu},p_{\nu}]=i_{eff}\hslash\delta^{\mu}_{\nu},
\end{equation}
while for a fermionic pair
\begin{equation}
 \{q^{\mu},q'^{\nu}\}=0; \quad \{q^{\mu},p_{\nu}\}=i_{eff}\hslash\delta^{\mu}_{\nu}.
\end{equation}
This leads to the desired non-commutativity amongst configuration variables and the corresponding momenta of  matter degrees of freedom, at the emergent level. Evidently, there is included now a 
operator time - energy commutation relation. In anticipation of the standard quantum theory that will eventually emerge from here, we identify the constant $\hbar$ with Planck's constant.

In ths sense, a Generalized Quantum Dynamics which does not refer to a classical time emerges from the underlying non-commutative special relativity in the statistical thermodynamic limit ~\cite{Lochan:2012}. One does have a concept of time-evolution, but this evolution is with respect to the proper time $s$ constructed from the trace of the operator spacetime line-element. In Section IV we will discuss how one possibly proceeds from this GQD to recover classical time. 

Furthermore, since at the fundamental matrix level, the theory is Lorentz invariant as shown in 
~\cite{Lochan-Singh:2011}, if we add another assumption of
 boundedness of $H_{eff}$ and existence of zero eigenvalue of $\vec{P}_{eff}$ corresponding to a unique eigenstate $\psi_0$, there exists a 
proposed correspondence between canonical ensemble average quantities and Lorentz-invariant Wightmann functions in the emergent field theory,
$$ \psi_0^{\dagger}\langle P({z_{eff}})\rangle_{\hat{AV}}\psi_0=\langle vac |P({{\cal X}_{eff}})|vac \rangle. $$

We can also obtain an equivalent Schr\"{o}dinger picture corresponding to the emergent Heisenberg picture of space-time dynamics. For that, we define
$$U_{eff}(s)=\exp{(-i_{eff}\hslash^{-1} s H_{eff})},$$ 
such that
$$\frac{d}{ds}U_{eff}(s)=-i_{eff}\hslash^{-1}H_{eff}U_{eff}(s).$$
Then, for a Heisenberg state vector $\psi$ we form Schr\"{o}dinger picture state vector $\psi_{schr}(s)$, for space-time degrees of freedom
$$ \psi_{schr}(s)=U_{eff}(s)\psi,$$
$$i_{eff}\hslash \frac{d}{ds}\psi_{schr}(s)=H_{eff}\psi_{schr}(s).$$
Thus we obtain Schr\"{o}dinger evolution for the phase-space variables at the canonical ensemble average level. 

We note that time and space continue to retain their operator status, although they now commute with each other.

\section{Trace dynamics and the quantum measurement problem}
Let us once again have a look at Fig. 4. We have thus far outlined how the lowermost arrow [NSR to GQD] is realized. In the next Section we will discuss the next arrow [GQD to the classical world]. For the purpose of the present section, let us assume the classical world as given: matter fields are classical and classical spacetime obeys the laws of general relativity. The universe is dominated by classical matter, which is responsible for the generation of a classical spacetime - in particular there exists a classical time with respect to which evolution can be defined. 

In such a classical world, how does one realize quantum theory, so essential to successfully describe the very large number of quantum phenomena observed in the laboratory? The traditional approach of course is to start from a classical dynamics for a system with given configuration varables and their canonical momenta, to replace Poisson brackets by commutation relations, hence introducing Planck's constant, and to replace Hamilton's equations of motion by Heisenberg equations of motion [equivalently the Schr\"{o}dinger equation].

This approach [and the equivalent path-integral formulation], although extremely successful, ought to be regarded as not completely satisfactory, and  `phenomenological' in nature. Because it pre-assumes as given the knowledge of its own limiting case, namely classical dynamics. One should not have to `quantize' a classical theory; rather there should be some guiding symmetry principles for developing a quantum theory, and then deriving classical mechanics from quantum theory as a limiting case. This requirement is in the same spirit whereby one does not arrive at special relativity by `relativizing' Galilean mechanics, or one does not arrive at general relativity by `general relativizing' Newtonian gravitation. The more fundamental theory stands on its own feet, and the limiting case only arises as an approximation - the prior knowledge of the limiting case should not be essential for the construction of the fundamental theory. 

An offshoot of arriving at quantum theory by `quantization' is that this leaves us without an understanding of the absence of macroscopic superpositions [the Schr\"{o}dinger cat paradox] and of the quantum measurement problem. [Unless of course one accepts the many-worlds interpretation as an explanation, or one believes in Bohmian mechanics as being the correct mathematical formulation of quantum theory].

Trace Dynamics ~\cite{Adler:04} sets out to derive quantum theory from an underlying matrix dynamics where select matter degrees of degrees $\hat{q}^i$ are described by non-commuting matrices [whereas the rest of the matter fields, which dominate the Universe, continue to be treated as classical] and a classical [Minkowski] spacetime is a given. These matrices represent bosonic / fermionic degrees of freedom, depending on whether they belong to the even / odd sector of the graded Grassmann algebra. Like in the previous section, a classical dynamics is constructed for these matrix degrees, with the difference that now time evolution is with respect to a classical time, as opposed to a proper time constructed from the operator spacetime line-element. Given a Trace Lagrangian, one derives Lagrange's equations of motion, a Hamiltonian, and Hamilton's equations of motion.  Once again, as a consequence of global unitary invariance there is a conserved charge with dimensions of action, the Adler-Millard charge
\begin{equation}
 \tilde{C} = \sum_{r\in B}[\hat{q}_r,\hat{p}_r] -\sum_{r\in F}\{\hat{q}_r,\hat{p}_r\},
\end{equation}
where the commutators are for bosonic degrees of freedom, and anticommutators are for fermionic degrees. 
This time round though, there is no pair such as $(E^i,t^i)$ in the commutators, because time is not an operator. In fact it should be emphasized that the construction in this section proceeds in very much the same fashion as in the previous section, except that a classical spacetime is given. More precisely, the approach adopted in the previous section was developed by us completely following the work of Adler and collaborators as described in this section.  This matrix dynamics is Lorentz invariant, under transformation of the ordinary space-time coordinates.

An equilibrium statistical mechanics for this matrix dynamics is constructed, as before, by maximizing the entropy, and as before it can be shown that the canonical average of $\tilde{C}$ takes the form
\begin{equation}
\langle\tilde{C}\rangle_{AV}=i_{eff}\hbar.
\end{equation}
A Ward identity holds, from which one deduces, after replacing the Adler-Millard charge by its canonical average, the standard quantum relations of quantum theory, the Heisenberg equations of motion, and by taking the non-relativistic limit one can write the equivalent description of the dynamics in terms of the Schr\"{o}dinger equation. The correspondence between canonical ensemble averages and Wightmann functions is proposed as before. In this way one recovers ordinary relativistic quantum field theory, and its non-relativistic limit,  from the underlying classical matrix dynamics. This is the step described by the lower arrow in the upper half of Fig. 4.

Something very remarkable is achieved next, by the upper arrow in the top half of Fig. 4. One examines the role played by the statistical fluctuations around equilibrium, for the case of the non-relativistic Schr\"{o}dinger equation.  These are taken into account by revisiting the Ward identity, and instead of replacing $\tilde{C}$ by its canonical average, one replaces $\tilde{C}$ by the canonical average plus correction terms. These correction terms represent the ever-present statistical fluctuations around equilibrium, analogous to the Brownian motion corrections to equilibrium thermodynamics. These fluctuations induce a [linear] modification of the non-relativistic Schr\"{o}dinger 
equation, the modifications being caused by the stochastic fluctuations, and if one assumes the fluctuations to be of the white noise type, they can be described by the It\^o representation of Brownian motion.

In order to make contact with the quantum measurement problem, one must now make a somewhat ad hoc assumption [which must eventually be justified from a deeper understanding of Trace Dynamics, and perhaps of the possible involvement of gravity]. The point is that the Schr\"{o}dinger equation, after including fluctuations, turns out not to be norm-preserving. Now one knows from particle number conservation in non-relativistic quantum theory that norm must be preserved during evolution. While norm-preservation must eventually be proved from deeper principles, for now one defines a new wave-function by dividing the original wave-function by its norm, so that the new wave-function preserves norm. This new wave-function obeys a {\it non-linear} Schr\"{o}dinger equation while continuing to depend on the statistical fluctuations.

This non-linear Schr\"{o}dinger equation contains within itself a special class, which coincides with the so-called models of Continuous Spontaneous Localization [CSL] developed by Ghirardi, Rimini, Weber and Pearle ~\cite{Ghirardi:86, Pearle:76, Ghirardi2:90,  Bassi:03} to explain the absence of macroscopic superpositions and to provide a dynamical explanation for the collapse of the wave-function during a quantum measurement. A prototype of such models is the one particle stochastic non-linear Schr\"{o}dinger equation
~\cite{Diosi:89}
\begin{equation} \label{eq:qmupl1}
d \psi_t  =  \left[ -\frac{i}{\hbar} H dt + \sqrt{\lambda} (q - \langle q \rangle_t) dW_t  
 - \frac{\lambda}{2} (q - \langle q \rangle_t)^2 dt \right] \psi_t,
\end{equation}
where $q$ is the position operator of the particle, $\langle q \rangle_t \equiv \langle \psi_t | q | \psi_t \rangle$ is the quantum expectation, and $W_t$ is a standard Wiener process which encodes the stochastic effect. Evidently, the stochastic term is nonlinear and also nonunitary. The collapse constant $\lambda$ sets the strength of the collapse mechanics, and it is chosen proportional to the mass $m$ of the particle according to the formula
$
\lambda = \frac{m}{m_0}\; \lambda_0,
$
where $m_0$ is the nucleon's mass and $\lambda_0$ measures the collapse strength.

This equation can be used to prove the absence of macroscopic superpositions and solve the quantum measurement problem, and furthermore its predictions for experiments in the mesoscopic regime differ from those of the standard linear Schr\"{o}dinger equation ~\cite{Bassi:03, RMP:2012, Essay:2012}. This allows the stochastic non-linear quantum quantum dynamics, and hence Trace Dynamics, albeit indirectly, to  be confirmed or ruled out by laboratory tests in the foreseeable future. The structure of the equation naturally provides an amplification mechanism - collapse becomes more and more important for larger systems. Furthermore, as can be anticipated by the very nature of its construction [norm-prservation], this non-linear equation dynamically reproduces the Born probability rule for the random outcomes of successive quantum measurements on an observable.

Although more remains to be done [why fluctuations should preserve norm; can the CSL model be uniquely derived from trace dynamics, is the collapse constant $\lambda$ a new constant of nature, or is it determined by already known fundamental constants via involvement of gravity in collapse], it is unquestionably true that trace dynamics provides a very natural and attractive avenue for understanding the origin of probabilities during quantum measurement, although the Schr\"{o}dinger dynamics is by itself deterministic. It has to do with the universal presence of statistical fluctuations: if the Schr\"{o}dinger equation is a thermodynamic approximation to the underlying matrix dynamics, the stochastic non-linear corrections to the Schr\"{o}dinger equation which are responsible for dynamical collapse, and the origin of probabilities, are a consequence of the unavoidable presence of fluctuations around thermodynamic equilibrium.

It should also be emphasized that the theory of wave-function collapse discussed here [CSL] is a non-relativistic theory, as also is the starting point wherein the connection between trace dynamics and CSL is developed. Despite several attempts, a relativisitic theory of wave-function collapse does not yet exist ~\cite{RMP:2012}. One clear difficulty is that the norm-preservation condition, which permits the construction of the non-linear stochastic Schr\"{o}dinger equation, is not necessarily available anymore.

\section{From the generalized quantum dynamics, to Trace Dynamics} 
The ideas discussed in this section are a report on work in progress, and hence have not yet taken final shape in terms of a mathematical formulation. 

Trace Dynamics takes a classical spacetime as given, and on this given spacetime it considers the matrix dynamics of selected degrees of freedom, for which quantum behaviour is derived. To our understanding, a fully consistent treatment of these select degrees, which is in accordance with the Einstein hole argument, should also associate an operator space-time with these degrees, as discussed in Section II. However, and this is crucial, one makes an {\it assumption} that this operator spacetime associated with these select degrees of freedom makes a very negligible impact on the classical spacetime produced by the dominant classical matter fields. This assumption is what allows one to proceed with a pre-given classical spacetime while developing trace dynamics. It is possible however, as discussed towards the end of this section, that this assumption may have to be revisited, in order to understand better the fundamental nature of EPR quantum correlations [no signalling, but yet an `action at a distance', as during the collapse of the wave-function].

One must face next the hard problem of understanding the transition from a GQD to a classical world. At a simplistic level, one could take the following approach. One should consider the statistical fluctuations about the equilibrium, at which GQD holds. However, one knows how to do that only in the non-relativistic case. The non-relativistic limit of the GQD cannot be defined by "going to speeds much less than speed of light", since time and space are still operators and there clearly is no classical notion of speed here. However, in the Lorentz transformations which define the invariance of the operator spacetime line-element, the one-parameter invariance  along a given direction is defined by the parameter $\beta$ which in the classical limit is defined as $v/c$. A non-relativistic limit of GQD can hence be defined by taking the limit $\beta\ll 1$. In this limit one can demand that the fluctuations preserve norm in the Schr\"{o}dinger equation, in which case the Schr\"{o}dinger equation is transformed to a non-linear equation, of which the CSL type stochastic equation is a special case. Evolution is described with respect to the proper time $s$ defined from the trace of the operator spacetime element, and the Hamiltonian depends on configuration degrees of freedom which include operator time. As before, one can consider the many-particle macroscopic limit and show that macroscopic superpositions are absent. However, something else extremely significant happens now. The absence of macroscopic superpositions in the matter sector implies the absence of superpositions of different spacetime quantum states corresponding to the operator status of space and time, thereby leading to the {\it emergence of a classical spacetime}. This is an important lesson, even though yet understood only in the non-relativistic and flat case: the emergence of a classical macroscopic description for matter comes hand in hand with the emergence of classical spacetime - the two are inseparable, and this inseparability is entirely in accord with the Einstein hole argument. If quantum theory is an emergent phenomenon [emerging from trace dynamics], so is classical spacetime an emergent phenomenon [emerging again from the generalized trace dynamics]. The matrix degrees of freedom may well be called the `atoms of spacetime'. 

A greater challenge is to understand the relativistic case: how is the ordinary spacetime of special relativity to be recovered from GQD, when the norm-preservation condition is not apparently available.

An even greater challenge is to recover classical gravity! When one proceeds from GQD to recover the classical world, not only should the classical spacetime manifold emerge, but there must emerge also classical gravity, which satisfies Einstein equations. Only then can consistency with the Einstein hole argument be ensured. Now GQD by itself has no gravity. Thus it seems we must return again to the lowermost level, and propose that gravity be introduced at the level of matrix dynamics itself, possibly by going from the `flat' operator spacetime element to the `curved' operator spacetime element:
\begin{equation}
ds^{2} = Trd\hat{s}^2\equiv Tr{\hat{g}_{\mu\nu}d\hat{x}^{\mu}d\hat{x}^{\nu}}.
\end{equation}

The expectation is that operator Einstein equations can be assumed to hold at the matrix dynamics level, and coarse graining would lead to Einstein equations for the canonically averaged operator metric, self-consistently coupled with the `curved space' GQD which depends on the canonically averaged operator metric. [While of course this idea remains to be developed mathematically, one cannot help noticing the resemblance it bears to the Schr\"{o}dinger-Newton system studied by Di\'{o}si ~\cite{Diosi:84} and 
Penrose ~\cite{Penrose:96} and others ~\cite{RMP:2012} in the context of studying gravity induced dynamical wavefunction collapse]. From here, one possibly proceeds to study the impact of statistical fluctuations on the equilibrium GQD and canonically averaged Einstein equations. This system is now non-linearly self-coupled, and it could be that one may not have to by hand bring in the assumption of norm-preservation to arrive at a stochastic non-linear CSL type collapse model which obeys the Born rule. In the macroscopic limit, such a non-linear system could be responsible for making both macroscopic objects and the associated spacetime and gravity behave classically. Once such a classical world is recovered, one can implement the construction described in Section III, for arriving at quantum theory starting from trace dynamics for the select degrees of freedom.

Our ideas may provide a useful way out for a better understanding of the apparent `action at a distance' which seems to prevail during the seemingly instantaneous collapse of the wave-function and in EPR-type quantum correlations. Perhaps one must not entirely disregard the implications of the operator space-time metric line-element associated with the [sub-dominant] quantum system, as was done in Section III while deriving quantum theory on a given classical space-time background. A quantum system always `carries' such a line-element with itself, in the sense that the most fundamental matrix level of description always exists, although we coarse grain it to arrive at what we observe at a higher level. Seen from the viewpoint of this operator line-element, which is non-commutative in nature, there is no point-structure to the spacetime associated with it, no definite light-cone structure, and no pre-given causal order, although it does have operator-level Lorentz invariance.  Thus from the point of view of this line-element, `wave-function collapse' can well happen in a unsurprising manner which otherwise appears as `instantaneous action at a distance' from the point of view of the externally given classical spacetime, because the latter possesses a causal structure. But this latter causal structure is not intrinsic to the quantum system under study - its something we choose to employ for our convenience, and then we `cry foul'! Indeed since there is no violation of special relativity in a EPR measurement, the apparent strangeness could simply be a case of trying to describe the process from an inaccurate perspective. Support for our idea also comes from an important recent paper ~\cite{Brukner:11}, where it has been shown that if one does not assume a predefined global causal order, there are multipartite quantum correlations which cannot be understood in terms of definite causal order and spacetime may emerge from a more fundamental structure in a quantum to classical transition.

In summary, in this work we have addressed the two key fundamental obstacles which still hold us back from getting a better understanding of quantum theory: the problem of time and the problem of quantum measurement. The problem of time suggests that a fundamental description of spacetime which is more compatible with quantum theory than the conventional one, is a non-commutative spacetime. The passage from a non-commutative spacetime to the commutative one that we see around us is through a coarse graining: akin to a passage from microscopic Newtonian mechanics to macroscopic thermodynamics via statistical mechanics. Quantum theory also emerges as the equilibrium description from the underlying level via a coarse graining. Statistical thermodynamics invariably implies Brownian motion fluctuations around equilibrium, and these are what result in quantum theory being an approximation to a stochastic non-linear theory, and dynamically explain the collapse of the wave-function and the emergence of probabilities during a quantum measurement. Thus the problem of time and the problem of quantum measurement are related to each other; their solution possibly springs from the same underlying source. Ongoing laboratory experiments are testing whether quantum theory is indeed an approximation to a non-linear theory, and these experiments also indirectly test the idea that the issues of time and measurement in quantum theory are related to each other.
\bigskip

\noindent{\bf Acknowledgements:} It is a pleasure to thank Angelo Bassi, Suratna Das, Kinjalk Lochan and Hendrik Ulbricht for collaboration and fruitful discussions. I would like to thank the organizers of the conference Quantum Malta 2012 for holding a very stimulating conference, and the conference participants for insightful discussions. I am grateful to Thomas Filk for illuminating conversations on quantum theory, and for encouraging me to write this article. I would also like to thank Albrecht von M\"{u}ller and the Parmenides Foundation for organizing the Parmenides Workshop: The present - perspectives from physics and philosophy (Wildbad Kreuth, Germany, October, 2006) where some early ideas leading to the present work were described ~\cite{Singh:2009}.

This work was made possible through the support of a grant
from the John Templeton Foundation. The opinions expressed in this publication are
those of the author and do not necessarily reflect the views of the John Templeton
Foundation. The support of the Foundational Questions Institute is also gratefully acknowledged.

\smallskip

A much more detailed bibliography of works relevant to this article can be found in ~\cite{RMP:2012}.

\newpage

\bibliography{biblioqmts3}

\end{document}